\documentclass[preprint,11pt,authoryear]{elsarticle}
\usepackage{lmodern}
\usepackage[margin=1.25in]{geometry}
\usepackage{lineno} 
\usepackage{setspace}
\usepackage{amsmath,amssymb,amsthm}
\usepackage{graphicx}
\usepackage{caption}
\usepackage{subcaption}
\usepackage{bbm}
\usepackage[mathscr]{eucal}
\usepackage{longtable}
\usepackage{multirow}
\usepackage{booktabs}
\usepackage{adjustbox}
\usepackage{url}
\usepackage{natbib}
\setcitestyle{aysep={}} 
\usepackage{soul}
\usepackage{color}
\usepackage[pdfpagemode=UseNone,bookmarksopen=false,colorlinks=true]{hyperref}
\hypersetup{linkcolor=blue, citecolor=blue,
  filecolor=blue, urlcolor=red} 

\newtheorem{theorem}{Theorem}[section]
\newtheorem{lemma}{Lemma}[section]
\newtheorem{assumption}{Assumption}
\newtheorem{definition}{Definition}
\newtheorem{remark}{Remark}

\newtheorem{proposition}{Proposition}[section]

\allowdisplaybreaks
 
\begin{document}

\begin{frontmatter}

\title{\textbf{Testing Cross-Sectional Dependence in Clustered Data}}

\author[affil1]{Subhodeep Dey}
\author[affil1]{Gopal K.\ Basak}
\author[affil2]{Samarjit Das\corref{cor1}}
\ead{samarjitd@gmail.com}

\address[affil1]{Theoretical Statistics and Mathematics Unit, Indian Statistical Institute, 203 B.T. Road, Kolkata 700108, West Bengal, India}
\address[affil2]{Economic Research Unit, Indian Statistical Institute, 203 B.T. Road, Kolkata 700108, West Bengal, India}

\cortext[cor1]{Corresponding author.}

\date{\today}

\begin{abstract}
In clustered data, cross-sectional dependence within clusters can be characterized as weak, semi-strong, or strong. This paper proposes a simple test to identify the specific nature of this dependence in linear regression models. By exploiting the variation in cluster sizes, our approach detects the underlying dependence structure in a computationally simple and robust manner. We develop a test statistic that is shown to be asymptotically chi-square distributed under the null hypothesis of weak dependence. Simulation studies and an empirical illustration using India's Household Consumption Expenditure Survey (HCES) data for 2022-23 demonstrate the usefulness of the proposed test.
\end{abstract}

 
\end{frontmatter}

\section{Introduction} \label{section: introduction}

Clustered data arise whenever observations are grouped, for example, students within schools or households within villages. Ignoring this within-cluster error correlation can severely understate standard errors \citep{moulton1990}. Consequently, a large literature has been developed for robust inference through appropriate correction of standard errors \citep[see][]{wooldridge2003cluster,cameron2008bootstrap,cameron2015,abadie2020sampling,canay2021wild,abadie2023,chiang2023genuinely,mackinnon2023guide}, and cluster-robust inference has become standard practice in empirical research. These methods have been extended to non-nested multiway clustering \citep{cameron2011robust,mackinnon2021wild}, the double/debiased machine learning framework \citep{chiang2022multiway} and dyadic data \citep{tabord2019inference}. The validity of cluster-robust inference, however, depends on the form of the within-cluster dependence. For example, under strong dependence, the conventional pooled ordinary least squares (POLS) estimator is inconsistent in the presence of dominant clusters \citep{hansen2019asymptotic, djogbenou2019, dey2024cross}. Also, unequal cluster sizes can distort inference even when the number of clusters $G$ is large \citep{carter2017,mackinnon2017wild}. Moreover, \citet{kojevnikov2023some} show that the variance of the POLS estimator cannot be consistently estimated in the presence of at least one large cluster. On the other hand, when the errors are uncorrelated within clusters, the cluster-robust correction is unnecessary. The heteroskedasticity-robust $t$-statistic of \citet{white1980heteroskedasticity} is then valid, whereas the cluster-robust variance estimator is less precise and yields tests with lower power \citep{cameron2015,abadie2023,mackinnon2023testing}. Therefore, detecting the presence and strength of within-cluster dependence is necessary for valid inference.

In the panel literature, dependence is classified as weak, semi-strong, or strong according to the growth rate of the largest eigenvalue of the error covariance matrix \citep{chudik2011weak, basak2018understanding}, and the exponent of this growth can be estimated from the data \citep{bailey2016exponent}. A substantial literature has developed formal diagnostic tests for cross-sectional dependence \citep[see][ for a survey]{sarafidis2012cross}. These include the classical Lagrange multiplier tests \citep{breusch1980lagrange, pesaran2008bias} and the widely used CD test \citep{pesaran2021general}, whose null hypothesis is weak dependence \citep{pesaran2015testing}. These tests are designed for panels with a time dimension, because they rely on pairwise correlation coefficients estimated over time. They cannot be applied directly to clustered cross-sectional data. For a single cross-section, spatial tests such as Moran's test \citep{moran1948interpretation, anselin1988spatial} are available, but they require the researcher to specify in advance which units are neighbors using a spatial weight matrix. A researcher working with clustered cross-sectional data, in which the within-cluster covariance structure is completely unrestricted, is therefore left without an analog of the CD test. Cluster-robust standard errors are computed, but whether the correction is needed and how severe the underlying dependence is, are rarely tested.

We propose a diagnostic test for cross-sectional dependence in linear regression models with clustered data, which exploits the variation in cluster sizes. When errors within a cluster are weakly dependent, the variance of the cluster's score, that is, the regressor-weighted sum of its errors, grows proportionately with the cluster size. When dependence is stronger, it grows faster. We therefore check whether the squared cluster-level score, scaled by cluster size, trends upward. This is implemented through a simple auxiliary regression on cluster sizes and a  robust test.

The rest of the paper is organized as follows. Section \ref{section: methodology} defines the three types of within-cluster dependence and constructs the test from an auxiliary regression of the normalized squared cluster scores on cluster sizes. It then derives the asymptotic properties of the test. Section \ref{section: simulation} reports the simulation results and an empirical application to household survey data. The simulations study the size and power of the test, and the application examines the dependence within states and districts. Section \ref{section: conclusion} concludes the paper.

\section{Methodology} \label{section: methodology}

\subsection{Model and dependence structure}

Consider a dataset of $G$ independent clusters, where cluster $g\in\{1,\dots,G\}$ contains $N_g$ observations. We consider one-way clustering only. The observation $Y_{gi}$ on the $i$-th individual of the $g$-th cluster is generated as
\[
Y_{gi} = X_{gi}'\theta + \epsilon_{gi}, \qquad i=1,\dots,N_g;\ g=1,\dots,G,
\]
with $X_{gi}$ a $k\times1$ vector of regressors. Stacking within clusters yields
\begin{equation} \label{eq: model}
Y_g = X_g\theta + \epsilon_g, \qquad g=1,\dots,G,
\end{equation}
where $Y_g = (Y_{g1}, \dots, Y_{gN_g})'$ is $N_g\times1$, $X_g = (X_{g1}', \dots, X_{gN_g}')'$ is $N_g\times k$, $\epsilon_g = (\epsilon_{g1}, \dots, \epsilon_{gN_g})'$ is $N_g\times1$, and $\theta$ is a $k\times1$ parameter vector.
Let $N_{tot}=\sum_{g=1}^G N_g$ be the total sample size and $\bar N=N_{tot}/G$ the average cluster size. The regressors may include an intercept. The parameter $\theta$ is estimated by pooled ordinary least squares (POLS),
\[
\hat\theta=\Big(\sum_{g=1}^G X_g'X_g\Big)^{-1}\sum_{g=1}^G X_g'Y_g,
\qquad e_g^*=Y_g-X_g\hat\theta .
\]

\begin{assumption}\label{assump: error independence and exogenous}
The sequence $\{(X_g,\epsilon_g)\}_{g=1}^G$ is independent across $g$, with $\mathbb{E}[\epsilon_g\mid X_g]=0$ and $\mathbb{E}[\epsilon_g\epsilon_g'\mid X_g]=\Omega_g$ positive definite. Moreover, $\lambda_{\min}(\Omega_g)\ge c$ for all $g$, for some constant $0<c<\infty$, and the diagonal elements of $\Omega_g$ are uniformly bounded.
\end{assumption}

\begin{assumption}\label{assump: design restrictions}
There exist finite constants $0<c_x\le C_x<\infty$ such that, almost surely, $c_xN_gI_k \preceq X_g'X_g \preceq C_xN_gI_k$ for every $g$, where `$\preceq$' denotes the Loewner order.
\end{assumption}

We now characterize the dependence structure of the errors, which is fundamental to our analysis.

\begin{definition}\label{def: dependence}
We define three kinds of cross-sectional dependence within cluster $g$.
\begin{enumerate}
\item \textbf{Strong dependence:} $\lambda_{\max}(\Omega_g)=O_{e}(N_g)$.\footnote{Here $z_n=O_{e}(a_n)$ means that $z_n$ is of the exact order of $a_n$, that is, $|z_n/a_n|$ is bounded above and bounded strictly away from zero.}
\item \textbf{Semi-strong dependence:} $\lambda_{\max}(\Omega_g)=O_{e}(h(N_g))$, where $h(N_g)\uparrow\infty$, and $\frac{h(N_g)}{N_g}\to0$, as $N_g\to\infty$.
\item \textbf{Weak dependence:} $\lambda_{\max}(\Omega_g)=O_{e}(1)$.
\end{enumerate}
\end{definition}

One example of strong dependence is that (almost) all individuals in the cluster are correlated or interconnected. One possible example of semi-strong dependence is that the number of dependent pairs increases with cluster size, but at a lower rate. Independence or finitely many dependent pairs are special cases of weak dependence. Weak dependence also covers dependence that decays sufficiently fast as observations become more distant according to some measure. By Definition \ref{def: dependence}, under weak dependence, there exist finite constants $0<c\le C<\infty$, such that $c\le\lambda_{\min}(\Omega_g)\le\lambda_{\max}(\Omega_g)\le C$. Further details and examples are described in \citet{basak2018understanding}. If cluster $g$ exhibits strong or semi-strong dependence, then $N_g$ must diverge along with $G\to\infty$. If instead $N_g$ stays finite while $G\to\infty$, we are always left with weakly dependent observations.

We now propose a test of the null hypothesis $H_0$ that the errors are weakly dependent within clusters, against the alternative hypothesis $H_1$ that the dependence is semi-strong or strong. More compactly, we test $H_0:\lambda_{\max}(\Omega_g)=O(1)$ for all $g$, against $H_1:\lambda_{\max}(\Omega_g)\to\infty$ as $N_g\to\infty$, for all $g$. Note that the strength of the dependence is characterized by the growth rate of $\lambda_{\max}(\Omega_g)$.
The eigenvalue $\lambda_{\max}(\Omega_g)$ is not estimable from a single realization of $\epsilon_g$.  Unlike $\Omega_g$, which is an $N_g\times N_g$ matrix whose dimension grows with the cluster size, the matrix $X_g'\Omega_gX_g$ is $k\times k$. Since $k$ is finite, its dimension does not grow with $N_g$, and it can be estimated from the cluster scores. For each covariate $j = 1, \ldots, k$, write $X_{g,j}$ for the $j$-th column of $X_g$, and
\[
q_{g,j}=X_{g,j}'\Omega_gX_{g,j},
\]
which is the variance of the cluster score $\sum_{i=1}^{N_g}X_{g,ij}\epsilon_{gi}$. The following assumption specifies the bounds of $q_{g,j}$ for different within-cluster dependence regimes.

\begin{assumption}\label{assump: rates}
For every $j=1,\dots,k$, the normalized cluster variance $q_{g,j}/N_g$ satisfies
\begin{enumerate}
\item $c_q\le q_{g,j}/N_g\le C_q$ for finite constants $0<c_q\le C_q<\infty$, that is, $q_{g,j}=O_{e}(N_g)$ under weak dependence,
\item $q_{g,j}/N_g=O_{e}(h(N_g))$, that is, $q_{g,j}=O_{e}(N_gh(N_g))$ under semi-strong dependence, 
\item $q_{g,j}/N_g=O_{e}(N_g)$, that is, $q_{g,j}=O_{e}(N_g^2)$ under strong dependence.
\end{enumerate}
\end{assumption}

Under weak dependence and Assumptions \ref{assump: error independence and exogenous} and \ref{assump: design restrictions}, part 1 holds automatically with $c_q=c\,c_x$ and $C_q=C\,C_x$, because $\|X_{g,j}\|^2=(X_g'X_g)_{jj}\in[c_xN_g,C_xN_g]$. 
In the standard cluster-robust literature, $q_{g,j}=X_{g,j}'\Omega_gX_{g,j}$ is of order $O(N_g)$ when the errors are uncorrelated within clusters, and of order $O(N_g^2)$ under strong within-cluster dependence, for example, equicorrelated or factor-type errors
\citep{cameron2015,mackinnon2019when,djogbenou2019,mackinnon2023guide}. 

\begin{remark}\label{rem: weak vs uncorrelated}
It is worth noting that the asymptotic inference based on the POLS estimator ($\hat\theta$)  is invariant to whether the errors exhibit weak dependence or $\Omega_g$ is diagonal for all g. Accordingly, the two settings may be treated interchangeably without loss of generality for all practical purposes.

\end{remark}

\begin{remark}\label{rem: dataset classification}
The three kinds of dependence in Definition \ref{def: dependence} are defined for an individual cluster, whereas the hypotheses concern the data as a whole. We therefore use the following convention. By strong dependence, we mean that a non-vanishing fraction, that is, $O_e(G)$, of the clusters exhibit strong dependence, while the remaining clusters may be semi-strongly or weakly dependent. This fraction need not be large, for example, $G/2$ or $G/3$. Similarly, semi-strong dependence of the data means that $O_e(G)$ clusters are semi-strongly dependent, no cluster is strongly dependent, although weakly dependent clusters may still exist. Weak dependence of the data means that all clusters are weakly dependent, with no strongly or semi-strongly dependent cluster. A cluster is called weakly, semi-strongly or strongly dependent according to the rate of $q_{g,j}$ in Assumption \ref{assump: rates}, with a cluster counted as non-weak if it is non-weak along at least one covariate.

This convention does not cover every configuration. If a vanishing fraction of the clusters, that is, $o(G)$ (for example a single cluster, or $\sqrt G$ clusters), is strongly or semi-strongly dependent while the rest are weakly dependent, the data are neither weakly dependent nor strongly or semi-strongly dependent in the above sense. The test is not designed to detect such sparse departures from weak dependence, and its behavior in this case is not covered by our results.
\end{remark}

\begin{remark} \label{rem: practical applicability of the assumptions}
    The dependence structure in data from the social sciences is not expected to vary so abruptly from cluster to cluster. Some degree of similarity is expected. The observations made in Remark \ref{rem: dataset classification} should be read as some pathological situations; they have only theoretical interest.  
\end{remark}

\subsection{Auxiliary regression and proposed test}

For each covariate $j=1,\dots,k$, define the normalized squared cluster score
\[
Z_{g,j}=\frac{\left(\sum_{i=1}^{N_g}X_{g,ij}e_{gi}^*\right)^2}{N_g},
\]
and stack these into the $k\times1$ vector $Z_g=(Z_{g,1},\dots,Z_{g,k})'$. By Lemma \ref{lemma: bias} below, $Z_{g,j}$ is an unbiased estimate of $q_{g,j}/N_g$ asymptotically. By Assumption \ref{assump: rates}, it therefore stays bounded under weak dependence and grows with $N_g$ under semi-strong or strong dependence. To evaluate cluster-size dependence across all $k$ covariates simultaneously, we regress the response vector $Z_g$ on the scalar cluster size $N_g$:
\begin{equation} \label{eq: auxiliary model with Xg}
    Z_g = \phi + \gamma N_g + \eta_g, \qquad g=1,\dots,G,
\end{equation}
where $\phi$ is a $k\times1$ intercept vector, $\gamma$ is the $k\times1$ vector of slopes representing the rate of scaling along each covariate, and $\eta_g$ is a $k\times1$ vector of auxiliary errors.

\begin{assumption}\label{assump: non dominance condition}
No single cluster dominates the sample size, i.e., $\displaystyle\lim_{G\to\infty}\max_{1\le g\le G}\frac{N_g}{N_{tot}}=0$.
\end{assumption}

\begin{lemma} \label{lemma: bias}
Under Assumptions~\ref{assump: error independence and exogenous},
\ref{assump: design restrictions} and \ref{assump: non dominance condition},
and weak dependence, for every $j=1,\dots,k$,
\[
\max_{1\le g\le G}\left|\mathbb{E}[Z_{g,j}\mid X]-\frac{X_{g,j}'\Omega_gX_{g,j}}{N_g}\right|\to0,\qquad\text{as }G\to\infty.
\]
\end{lemma}

Testing whether $Z_g$ scales with $N_g$ is equivalent to testing $H_0^{(j)}:\gamma_j=0$, $j=1,\ldots,k$. Combining these restrictions, the null hypothesis $H_0$ of weak dependence can be restated as $H_0:\gamma=\mathbf 0$, and the alternative hypothesis as $H_1:\gamma\neq\mathbf 0$. Under $H_1$, $O(G)$ clusters are semi-strongly or strongly dependent.\footnote{For all practical purposes, as in Remark \ref{rem: practical applicability of the assumptions}, $O(G)$ should be read as all $G$ clusters.} Hence, a rejection of $H_0$ indicates the presence of non-weak dependence.

The dependence induced by $\hat\theta$ is asymptotically negligible under Assumption~\ref{assump: non dominance condition}, since no single cluster dominates the pooled estimator. The auxiliary regressor $N_g$ is non-stochastic. Therefore, no conditioning on design is required, and we state the assumptions on $\eta_g$ unconditionally.

\begin{assumption}\label{assump: auxiliary error mean 0 with Xg}
The auxiliary errors $\eta_1,\dots,\eta_G$ are independent across $g$ with $\mathbb{E}[\eta_g]=0$.
\end{assumption}

\begin{assumption}\label{assump: auxiliary error moments with Xg}
Let $\Sigma_g=\mathrm{Var}(\eta_g)$. Under weak dependence, there exist finite constants $\sigma_{\min}^2>0$ and $\kappa<\infty$ such that, uniformly in $g$, $\lambda_{\min}(\Sigma_g)\ge\sigma_{\min}^2$. Also, for every unit vector $\lambda\in\mathbb R^k$, $\mathbb{E}\big[(\lambda'\eta_g)^4\big]\le\kappa$.
\end{assumption}

The fourth-moment bound controls the variance of the heteroskedasticity-consistent covariance matrix (HCCM) estimator introduced below. By the Cauchy-Schwarz inequality, Assumption~\ref{assump: auxiliary error moments with Xg} also gives $\lambda_{\max}(\Sigma_g)\le\sqrt\kappa=\sigma^2_{\max}$, say. Applying OLS to \eqref{eq: auxiliary model with Xg} gives
\begin{equation}
    \hat\gamma=\frac{\sum_{g=1}^G(N_g-\bar N)Z_g}{\sum_{g=1}^G(N_g-\bar N)^2},\qquad
    \hat\phi=\bar Z-\hat\gamma\bar N,\qquad \bar Z=\frac1G\sum_{g=1}^GZ_g.
\end{equation}
Note that $\hat\gamma$ further reduces to $\hat\gamma=S_G\big/\sum_{g=1}^G(N_g-\bar N)^2$ with $S_G=\sum_{g=1}^G(N_g-\bar N)Z_g$. Substituting \eqref{eq: auxiliary model with Xg}, and using  $\sum_{g=1}^G(N_g-\bar N)N_g=\sum_{g=1}^G(N_g-\bar N)^2$,
\begin{equation}\label{eq: S_G decomposition}
    S_G=\gamma\sum_{g=1}^G(N_g-\bar N)^2+\sum_{g=1}^G(N_g-\bar N)\eta_g .
\end{equation}
Under $H_0:\gamma=\mathbf{0}$, $S_G=\sum_{g=1}^G(N_g-\bar N)\eta_g$, so $\hat\gamma$ is driven by the multivariate score vector $S_G$ alone and does not depend on $\phi$. By Assumption~\ref{assump: auxiliary error mean 0 with Xg}, under $H_0$, the score $S_G$ has mean zero and variance $V_G=\sum_{g=1}^G(N_g-\bar N)^2\Sigma_g$, where $\Sigma_g=\mathrm{Var}(\eta_g)$ is as in Assumption~\ref{assump: auxiliary error moments with Xg}. The lower bound in that assumption makes $V_G$ invertible, since $V_G\succeq\sigma_{\min}^2\sum_{g=1}^G(N_g-\bar N)^2I_k$. Because the $k$ covariates are generally correlated, we estimate $V_G$ by the HCCM estimator of \citet{white1980heteroskedasticity} (HC0). The residual vector for the $g$-th cluster is $\hat\eta_g=Z_g-\hat\phi-\hat\gamma N_g$, and the HCCM estimator of the covariance matrix of $\hat\gamma$ is 
\begin{equation}
    \widehat{\mathrm{Var}}_{rob}(\hat\gamma)=\frac{\hat V_G}{\big(\sum_{g=1}^G(N_g-\bar N)^2\big)^2},\qquad
    \hat V_G=\sum_{g=1}^G(N_g-\bar N)^2\hat\eta_g\hat\eta_g'.
\end{equation}
Here, the proposed test statistic is
\begin{equation}
    W=\hat\gamma'\big[\widehat{\mathrm{Var}}_{rob}(\hat\gamma)\big]^{-1}\hat\gamma=S_G'\hat V_G^{-1}S_G,
\end{equation}
and the marginal $t$-statistic for $H_0^{(j)}$ is $t_j=\hat\gamma_j/\sqrt{\widehat{\mathrm{Var}}_{rob}(\hat\gamma)_{jj}}$ for every $j=1,2,\dots,k$.

\begin{remark} \label{rem: role of bias lemma}
Lemma~\ref{lemma: bias} shows that, under weak dependence, $\mathbb{E}[Z_{g,j}\mid X]$ equals $q_{g,j}/N_g$ up to a bias that vanishes uniformly in $g$. A uniformly vanishing bias is not automatically negligible for the test. Let us denote this bias as $b_g = (b_{g,1}, \ldots, b_{g,k})'$ with $b_{g,j} = \mathbb{E}[Z_{g,j}\mid X] - q_{g,j}/N_g$ for $j = 1, \ldots, k$. By \eqref{eq: explicit bias bound} from \ref{appendix: proofs}, $|b_{g,j}|\le K_0N_g/N_{tot}$ for a finite constant $K_0$. The bias is thus larger for larger clusters, so it moves together with $N_g$, the regressor in the auxiliary regression, and it could create a spurious slope in $N_g$ that does not average out across clusters. It is nevertheless negligible. The bias enters the slope only through the term $\sum_{g=1}^G(N_g - \bar N)b_g$ in $S_G$. By the triangle inequality, the Cauchy--Schwarz inequality and $\sum_gN_g^2\le N_{tot}\max_gN_g$, for each $j$,
\[
  \left|\sum_{g=1}^G(N_g - \bar N)b_{g,j}\right|
  \le\frac{K_0}{N_{tot}}\sum_{g=1}^G|N_g - \bar N|N_g
  \le K_0\left(\sum_{g=1}^G(N_g - \bar N)^2\right)^{1/2}\left(\frac{\max_gN_g}{N_{tot}}\right)^{1/2}.
\]
The statistic uses $S_G$ standardized by $V_G^{-1/2}$, where $V_G\succeq\sigma_{\min}^2\sum_g(N_g - \bar N)^2I_k$ by Assumption~\ref{assump: auxiliary error moments with Xg}. Dividing by $\sigma_{\min}\left(\sum_g(N_g - \bar N)^2\right)^{1/2}$, the total contribution of the bias to the standardized numerator of $\hat\gamma$ is at most $(K/\sigma_{\min})\sqrt{\max_gN_g/N_{tot}}$ for a finite constant $K$. This quantity tends to zero by Assumption~\ref{assump: non dominance condition}. Therefore,  the bias from estimating $\beta$ does not affect the null distribution of the test statistic, which is why the mean-zero condition on $\eta_g$ in Assumption~\ref{assump: auxiliary error mean 0 with Xg} is reasonable under $H_0$. The same estimation step makes the $\eta_g$ weakly dependent across clusters, which is why independence remains an assumption.
\end{remark}


\begin{assumption}\label{assump: nondegenerate sizes}
There exists $v>0$ such that, for all large $G$, $\frac1G\sum_{g=1}^G(N_g-\bar N)^2\ge v$.
\end{assumption}

\begin{assumption}\label{assump: no dominance variation}
No single cluster dominates the variation in cluster sizes, i.e.,
\[
\lim_{G\to\infty}\frac{\max_{1\le g\le G}(N_g-\bar N)^2}{\sum_{g=1}^G(N_g-\bar N)^2}=0.
\]
\end{assumption}

Assumptions \ref{assump: nondegenerate sizes} requires the total variation in cluster sizes, $\sum_{g=1}^G(N_g-\bar N)^2$, to be at least of the order of $G$, so that the slope in the auxiliary regression is identified. Assumption \ref{assump: no dominance variation} requires that this variation is spread over many clusters rather than carried by a few, which gives asymptotic normality. The assumptions hold whenever $\max_{1\le g\le G}(N_g-\bar N)^2/G\to0$. This covers bounded cluster sizes as well as sizes whose largest deviation grows more slowly than $\sqrt G$.

\begin{remark}\label{rem: sufficient condition}
Assumption \ref{assump: nondegenerate sizes} yields a sufficient condition for Assumption \ref{assump: no dominance variation}. If $\max_{1\le g\le G}(N_g-\bar N)^2/G\to0$, then the ratio in Assumption \ref{assump: no dominance variation} is at most $v^{-1}\max_g(N_g-\bar N)^2/G\to0$. This holds in particular, if $\max_g(N_g-\bar N)^2<\infty$, for example, when $N_g\in\{N_{\min},\dots,N_{\max}\}$ with fixed $N_{\min},N_{\max}$. 
\end{remark}

\begin{theorem}\label{thm: wald}
Under Assumptions~\ref{assump: auxiliary error mean 0 with Xg}--\ref{assump: no dominance variation} and $H_0:\gamma=\mathbf 0$, 
\[
W=S_G'\hat V_G^{-1}S_G\;\xrightarrow{d}\;\chi^2_{(k)},\qquad G\to\infty.
\]
\end{theorem}

The proof combines Lemmas \ref{lemma: CLT} and \ref{lemma: variance} in \ref{appendix: proofs}. Lemma \ref{lemma: CLT} provides a central limit theorem for the score $S_G$. Lemma \ref{lemma: variance} gives the consistency of the HCCM estimator $\hat V_G$. Slutsky's theorem and the continuous mapping theorem gives the Theorem \ref{thm: wald}. The formal proof of this theorem is provided in \ref{appendix: proofs}.

We now show that the test is consistent, i.e., if $\gamma\neq\mathbf 0$, the statistic diverges and $H_0$ is rejected with probability approaching one. Under the alternative, the matrices $\Sigma_g=\mathrm{Var}(\eta_g)$ need no longer be bounded. We therefore drop Assumption \ref{assump: auxiliary error moments with Xg} and add the following two conditions, which allow $\Sigma_g$ to diverge as $G\to\infty$.

\begin{assumption} \label{assump: Sigma_g diveregence with Xg}
As $G\to\infty$, $\displaystyle\frac{\max_{1\le g\le G}\lambda_{\max}(\Sigma_g)}{\sum_{g=1}^G(N_g-\bar N)^2}\to0$.
\end{assumption}

\begin{remark}\label{rem: Sigma growth}
A sufficient condition for Assumption \ref{assump: Sigma_g diveregence with Xg} is $\lambda_{\max}(\Sigma_g)\le C N_{\max}^{2r}$ for constants $C<\infty$ and $r\ge0$, where $N_{\max}=\max_gN_g$, together with $\sqrt{\sum_{g=1}^G(N_g-\bar N)^2}\,/\,N_{\max}^r\to\infty$. Then, the ratio in Assumption \ref{assump: Sigma_g diveregence with Xg} is at most $C\big(N_{\max}^{r}/\sqrt{\sum_{g=1}^G(N_g-\bar N)^2}\big)^2\to0$. This allows $\Sigma_g$ to diverge with cluster size.
\end{remark}

\begin{assumption} \label{assump: eta_bar variance condition with Xg}
There exists a finite constant $K_1$ such that, for all sufficiently large $G$,
\[
\frac{\sum_{g=1}^G(N_g-\bar N)^2}{G^2}\sum_{g=1}^G\Sigma_g\;\preceq\;K_1V_G.
\]
\end{assumption}

\begin{proposition} \label{prop:consistency_mult}
Suppose that the auxiliary regression \eqref{eq: auxiliary model with Xg} holds under $H_1: \gamma\neq \mathbf 0$. Under Assumptions~\ref{assump: auxiliary error mean 0 with Xg} and \ref{assump: nondegenerate sizes}--\ref{assump: eta_bar variance condition with Xg}, $\mathbb{P}(W>\delta)\to1$, as $G\to\infty$, for every fixed $\delta>0$.
\end{proposition}

Proposition~\ref{prop:consistency_mult} reveals that the test rejects whenever the normalized score has a non-zero linear trend in the cluster size. The proof of this proposition is given in \ref{appendix: proofs}. 

\begin{remark} \label{rem: intercept}
Consider the special case $X_g=\mathbf 1_{N_g}$, i.e., $k=1$, $\theta=\mu$ and $Y_g=\mathbf 1_{N_g}\mu+\epsilon_g$. Here $\hat\theta$ is the grand mean, $Z_g=\zeta_g/N_g$ with $\zeta_g=\big(\sum_{i=1}^{N_g}e^*_{gi}\big)^2$, and the auxiliary regression becomes $\zeta_g/N_g=\alpha+\beta N_g+\eta_g$ with $H_0:\beta=0$. The Wald statistic reduces to $W=T^2$, where $T$ is the two-sided $t$-statistic based on the HCCM estimator, so the test is the usual robust $t$-test and $T\xrightarrow{d}N(0,1)$ under $H_0$. Here, $\mathbf 1_{N_g}'\Omega_g\mathbf 1_{N_g}$ is the variance of the cluster sum of the outcome, so the type of dependence is detected directly from $\Omega_g$.
\end{remark}

\begin{remark}  \label{rem: semi-strong}
The linear auxiliary regression is designed to detect strong dependence, since $\mathbb{E}[Z_{g,j}]$ grows linearly in $N_g$. To capture general dependence, suppose that $q_{g,j}/N_g$ is of exact order $N_g^{\beta_j}$ with an exponent $\beta_j\in[0,1]$ that is common to all clusters. Then $\beta_j=0$ corresponds to weak dependence, $0<\beta_j<1$ to semi-strong dependence, and $\beta_j=1$ to strong dependence. Now, let us consider the following auxiliary regression for each covariate $j$:
\[
\log Z_{g,j}=\alpha_j+\beta_j\log N_g+\xi_{g,j},\qquad g=1,\dots,G.
\]
We estimate $\beta_j$ by OLS with an HCCM standard error, as in the linear auxiliary regression. The procedure has two steps. In the first step, we test $H_0:\beta_j=0$ against $H_1:\beta_j>0$ with a one-sided $t$-test, which checks whether the dependence is weak. If $H_0$ is rejected, then in the second step we test $H_{10}:\beta_j=1$ against $H_{11}:\beta_j<1$ with a one-sided $t$-test, which checks whether the dependence is strong rather than semi-strong. If the first step does not reject, we conclude that the dependence is weak. If the first step rejects and the second does not, we conclude that the dependence is strong. Finally, if both the steps reject, we conclude that the dependence is semi-strong. 
\end{remark}

\begin{remark}
 It is not difficult to construct a few pathological examples where $X_{g,j}'\Omega_gX_{g,j}$ is of order $O_P(N_g)$ even under strong dependence in the errors. In such cases, the test has no power for the $j$-th regressor. However, the joint test still may have power.
\end{remark}

\begin{remark}\label{rem: superblock}
As an alternative to $\hat\theta$, following \citet{dey2026linear}, the $G$ clusters may be partitioned into $D$ disjoint superblocks, with $S_l$ the set of cluster indices in superblock $l$ and $P_l=|S_l|$ its number of clusters. Two alternatives to the pooled estimator are the superblock average of cluster-specific estimates, $\hat\theta_{avg,l}=P_l^{-1}\sum_{g\in S_l}\hat\theta_g$, and the superblock OLS estimator, $\hat\theta_{ols,l}=\big(\sum_{g\in S_l}X_g'X_g\big)^{-1}\sum_{g\in S_l}X_g'Y_g$, with residuals computed from the estimate of the superblock containing the cluster. The argument in the proof of Lemma \ref{lemma: bias} applies with the pooled design matrix replaced by its superblock analog, so that the bias bound involves $N_g/\sum_{h\in S_l}N_h$ instead of $N_g/N_{tot}$. This ratio must vanish uniformly in $g$, which requires $\min_lP_l\to\infty$. Assumption \ref{assump: non dominance condition} is then replaced by the corresponding condition within superblocks, and the remaining results are unchanged.
\end{remark}

In the next section, we evaluate the finite-sample performance of the proposed test in simulations, and provide an application to household survey data.

\section{Simulation Study and Empirical Application} \label{section: simulation}

\subsection{Simulation Study}

We first study the finite-sample size and power of the proposed test in the intercept-only model of Remark~\ref{rem: intercept}, that is, model \eqref{eq: model} with $k=1$ and $X_g=\mathbf 1_{N_g}$. Here, $Y_{gi}=\mu+\epsilon_{gi}$ with true value $\mu=2$. The parameter $\mu$ is estimated by the grand mean, and $Z_g$ are formed as in Section~\ref{section: methodology}. We test $H_0:\gamma=0$ in the auxiliary regression \eqref{eq: auxiliary model with Xg}, comparing the Wald statistic $W=T^2$ with the $\chi^2_1$ distribution at the nominal $5\%$ level. The cluster sizes $N_g$ are drawn from the discrete uniform distribution on $\{50,\ldots,120\}$. The cluster sizes and the covariance matrices $\Omega_g$ are held fixed across replications. We vary the number of clusters $G$ and use $R = 10{,}000$ replications per design.

Two weak-dependence designs are used to assess size. In design WD1, $\Omega_g$ is diagonal with diagonal elements $\sigma_{gi}^2\sim U(0.5,4.5)$, so the errors are independent but heteroskedastic. In design WD2, $\Omega_g$ is a band matrix with bandwidth 2. Its off-diagonal entries are drawn from $U(5,10)$, and each diagonal entry is the maximum of a $U(10,20)$ draw and $1.05$ times the absolute row sum.  This ensures positive definiteness and a bounded maximum eigenvalue for $\Omega_g$. Two strongly dependent designs are used to assess power. In design SD1, $\Omega_g=N_g\,M_gM_g'/\lambda_{\max}(M_gM_g')$, where $M_g$ is an $N_g\times N_g$ matrix with i.i.d.\ $U(-5,10)$ entries, so that $\lambda_{\max}(\Omega_g)=N_g$. In design SD2, the errors follow a common factor model, $\Omega_g=\lambda_g\lambda_g'+\Psi_g$, with loadings $\lambda_{gi}\sim U(0.5,1.5)$ and a diagonal $\Psi_g$ with entries $U(0.5,4.5)$, so that $\Omega_g$ becomes a strongly dependent matrix.

\begin{figure}[htbp]
\centering
\begin{subfigure}[t]{0.48\textwidth}
  \centering
  \caption{Size: WD1 and WD2}
  \includegraphics[width=\textwidth]{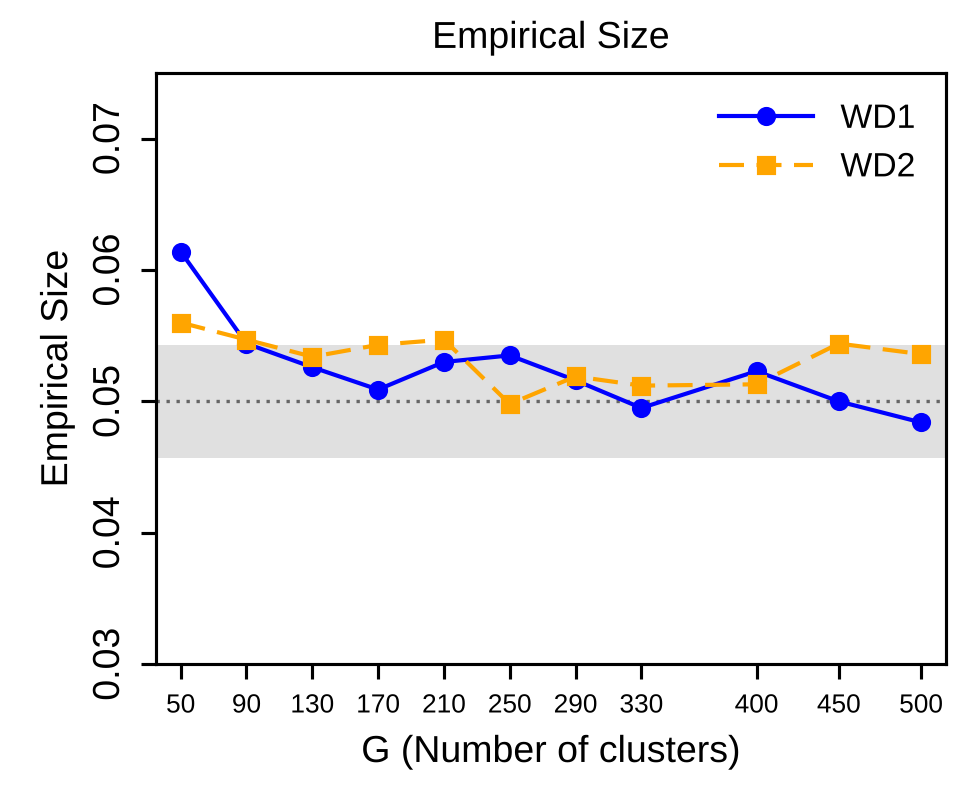}
  \label{fig:dep-size-k1}
\end{subfigure}\hfill
\begin{subfigure}[t]{0.48\textwidth}
  \centering
  \caption{Power: SD1 and SD2}
  \includegraphics[width=\textwidth]{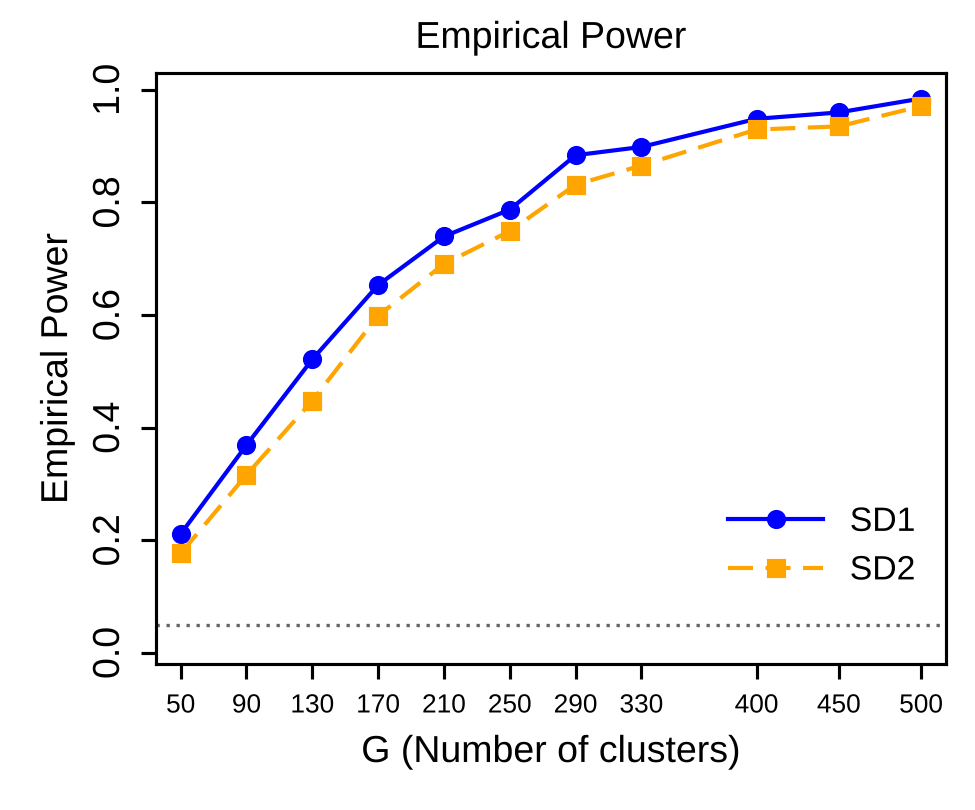}
  \label{fig:dep-power-k1}
\end{subfigure}
\caption[Empirical size and power of the proposed dependence test in the intercept-only model.]{Empirical size and power of the proposed dependence test in the intercept-only model; $k=1$. Cluster sizes drawn from the discrete uniform distribution on $\{50,\ldots,120\}$ and $R=10{,}000$. (a) Size under the weak dependent designs WD1 and WD2, with the shaded $95\%$ Monte Carlo interval. (b) Power under the strong dependent designs SD1 and SD2.}
\label{fig:dep-rejection-k1}
\end{figure}

\autoref{tab:dep-sim-k1} and \autoref{fig:dep-rejection-k1} report the rejection frequencies. The shaded band in \autoref{fig:dep-size-k1} is the $95\%$ Monte Carlo interval $0.05\pm z_{0.975}\sqrt{0.05\times0.95/R}\simeq[0.046,0.054]$.

\begin{table}[htbp]
\centering
\caption{Empirical size and power of the proposed dependence test in the intercept-only model.}
\label{tab:dep-sim-k1}
\begin{tabular}{c cc cc}
\hline \hline
 & \multicolumn{2}{c}{Size} & \multicolumn{2}{c}{Power} \\
\cline{2-3} \cline{4-5}
$G$ & WD1 & WD2 & SD1 & SD2 \\
\hline
50  & 0.0614 & 0.0560 & 0.2113 & 0.1767 \\
90  & 0.0544 & 0.0547 & 0.3691 & 0.3163 \\
130 & 0.0526 & 0.0534 & 0.5213 & 0.4477 \\
170 & 0.0509 & 0.0543 & 0.6533 & 0.5989 \\
210 & 0.0530 & 0.0547 & 0.7395 & 0.6906 \\
250 & 0.0535 & 0.0498 & 0.7872 & 0.7490 \\
290 & 0.0516 & 0.0519 & 0.8838 & 0.8315 \\
330 & 0.0495 & 0.0512 & 0.8985 & 0.8655 \\
400 & 0.0523 & 0.0513 & 0.9485 & 0.9295 \\
450 & 0.0500 & 0.0544 & 0.9598 & 0.9350 \\
500 & 0.0484 & 0.0536 & 0.9845 & 0.9709 \\
\hline \hline
\end{tabular}

\vspace{1.0ex}
\begin{minipage}{0.9\textwidth}
\footnotesize \textit{Note:} The number of replications is $10{,}000$; $k=1$. Cluster sizes are drawn from the discrete uniform distribution on $\{50,\dots,120\}$. The nominal level is $0.05$. Size designs: WD1 (independent heteroskedastic) and WD2 (banded). Power designs: SD1 (scaled $MM'$) and SD2 (factor model).
\end{minipage}
\end{table}

For the size evaluation, we compare the rejection frequencies under two types of weak dependence with the $5\%$ level of significance. With $50$ clusters, the empirical size in both designs is slightly above the nominal level. As $G$ increases, the empirical size for both the designs reaches the nominal level, (\autoref{tab:dep-sim-k1} and \autoref{fig:dep-size-k1}). This supports the asymptotic result of Theorem~\ref{thm: wald}.
Under strong dependence, the empirical power increases with $G$ in both designs (Table~\ref{tab:dep-sim-k1} and Figure~\ref{fig:dep-power-k1}) and approaches one for large $G$. Since $\lambda_{\max}(\Omega_g)$ is of order $N_g$ in both designs, the power is similar in the two designs, with slightly higher power in design SD1. In both designs, the power increases with $G$, in line with Proposition~\ref{prop:consistency_mult}.

\begin{figure}[htbp]
\centering
\begin{subfigure}[t]{0.48\textwidth}
  \centering
  \caption{Size: WD1 and WD2}
  \includegraphics[width=\textwidth]{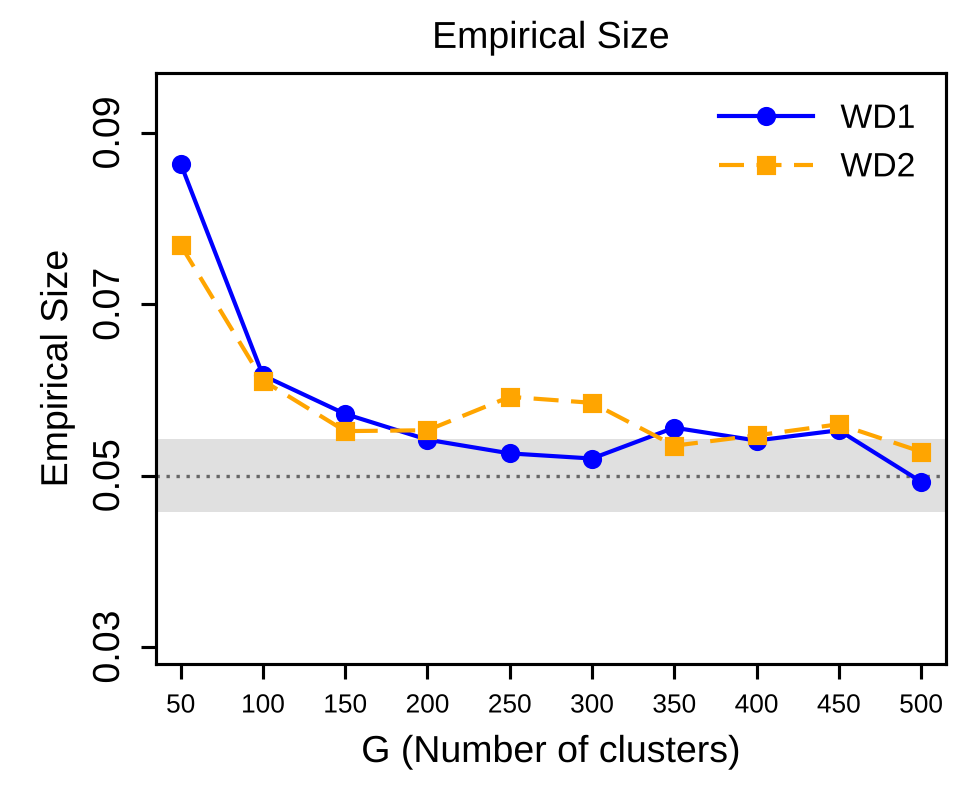}
  \label{fig:dep-size-k3}
\end{subfigure}\hfill
\begin{subfigure}[t]{0.48\textwidth}
  \centering
  \caption{Power: SD1 and SD2}
  \includegraphics[width=\textwidth]{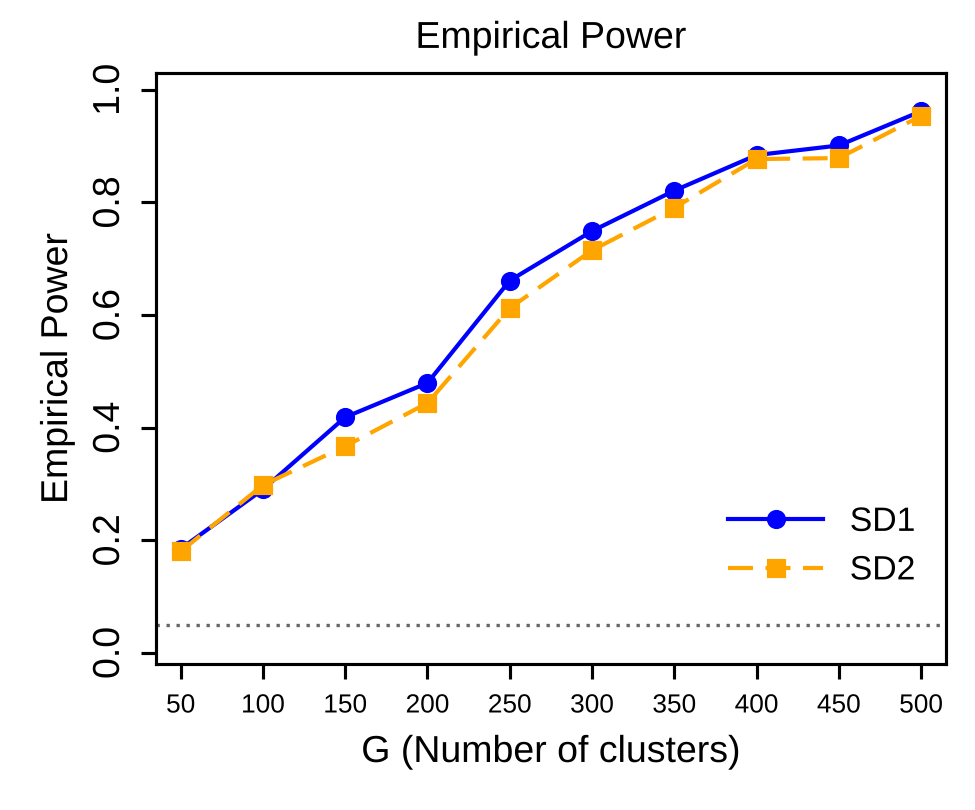}
  \label{fig:dep-power-k3}
\end{subfigure}
\caption[Empirical size and power of the proposed dependence test.]{Empirical size and power of the proposed dependence test in the regression model with $k=3$. Cluster sizes drawn from the discrete uniform distribution on $\{50,\ldots,120\}$ and $R=10{,}000$. (a) Size under the weak dependent designs WD1 and WD2, with the shaded $95\%$ Monte Carlo interval. (b) Power under the strong dependent designs SD1 and SD2.}
\label{fig:dep-rejection-k3}
\end{figure}

We now generalize the design to the regression model \eqref{eq: model} with $k=3$ regressors: an intercept and two independent standard normal covariates. Here, the true value is $\beta=(2,1,0.5)'$. The parameter $\beta$ is estimated by POLS, $\hat\beta_{POLS}$, and we compare  $W$ with the $\chi^2_3$ distribution at the nominal $5\%$ level. The covariates are held fixed across replications. The cluster sizes, the four covariance designs and the number of replications are the same as in the intercept-only model. \autoref{tab:dep-sim-k3} and \autoref{fig:dep-rejection-k3} report the rejection frequencies for different values of $G$.

\begin{table}[htbp]
\centering
\caption{Empirical size and power of the proposed dependence test.}
\label{tab:dep-sim-k3}
\begin{tabular}{c cc cc}
\hline \hline
 & \multicolumn{2}{c}{Size} & \multicolumn{2}{c}{Power} \\
\cline{2-3} \cline{4-5}
$G$ & WD1 & WD2 & SD1 & SD2 \\
\hline
50  & 0.0864 & 0.0769 & 0.1849 & 0.1803 \\
100 & 0.0617 & 0.0611 & 0.2903 & 0.2980 \\
150 & 0.0572 & 0.0552 & 0.4186 & 0.3671 \\
200 & 0.0542 & 0.0553 & 0.4794 & 0.4438 \\
250 & 0.0526 & 0.0592 & 0.6612 & 0.6123 \\
300 & 0.0520 & 0.0585 & 0.7491 & 0.7149 \\
350 & 0.0556 & 0.0535 & 0.8209 & 0.7906 \\
400 & 0.0541 & 0.0547 & 0.8841 & 0.8770 \\
450 & 0.0553 & 0.0560 & 0.9011 & 0.8787 \\
500 & 0.0493 & 0.0528 & 0.9619 & 0.9536 \\
\hline \hline
\end{tabular}

\vspace{1.0ex}
\begin{minipage}{0.9\textwidth}
\footnotesize \textit{Note:} The number of replications is $10{,}000$; $k=3$. Cluster sizes are drawn from the discrete uniform distribution on $\{50,\dots,120\}$. The nominal level is $0.05$. Size designs: WD1 (independent heteroskedastic) and WD2 (banded). Power designs: SD1 (scaled $MM'$) and SD2 (factor model).
\end{minipage}
\end{table}

With $k=3$, the size distortion for a small number of clusters is larger than in the intercept-only model, since the HCCM estimator now estimates a $3\times3$ matrix from the same number of clusters. As $G$ grows, the size moves close to the nominal level in both designs (\autoref{tab:dep-sim-k3} and \autoref{fig:dep-size-k3}).
Under strong dependence, the empirical power increases with $G$ in both designs and approaches one for large $G$ (Table~\ref{tab:dep-sim-k3} and Figure~\ref{fig:dep-power-k3}), in line with Proposition~\ref{prop:consistency_mult}.

\subsection{Empirical Illustration}
 
We apply the proposed test to household expenditure data from the Household Consumption Expenditure Survey (HCES) 2022--23 of India. The sample contains $244{,}884$ households, all with positive consumption expenditure. The outcome is the logarithm of monthly per capita consumption expenditure (MPCE), computed as the total value of household consumption over all items divided by household size. The regression model is
\begin{equation*}
\log(\mathrm{MPCE}_{gi}) = \beta_0 + \beta_1 \log(\mathrm{hhsize}_{gi}) + \beta_2\, \mathrm{urban}_{gi} + \beta_3\, \mathrm{childshare}_{gi} + \varepsilon_{gi},
\end{equation*}
where $\mathrm{urban}_{gi}$ equals one for urban households and $\mathrm{childshare}_{gi}$ is the share of children in the $i$-th household and $g$-th cluster. We consider two clustering levels: state and district. Clusters with fewer than five households are dropped. For states as clusters, $G=36$ with cluster size between $512$ and $28{,}772$ households. The district level clustering has $G=692$ clusters with cluster size between $5$ and $2{,}057$ households. Here, $k=4$ and the test statistic has four degrees of freedom. 
  
Table \ref{tab:hces_wald} reports the proposed test of testing $H_0\colon \gamma = 0$ in the auxiliary regression $Z_g = \phi + \gamma N_g + \eta_g$, together with the covariate-specific $t$-statistics of $\hat{\gamma}_j$. 
 
\begin{table}[htbp]
\centering
\caption{Dependence test for HCES India 2022--23.}\label{tab:hces_wald}
\begin{tabular}{lrrrrrrr}
\toprule
 & & & & \multicolumn{4}{c}{$t$-statistic for $\hat\gamma_j$} \\
\cmidrule(lr){5-8}
Clustering level & $G$ & $W$ & $p$-value & const & $\log(\mathrm{hhsize})$ & urban & childshare \\
\midrule
State    & $36$       & $11.34$  & $0.023$              & $1.34$ & $2.07$ & $0.63$  & $1.99$ \\
District & $692$      & $37.81$  & $1.2\times10^{-7}$   & $5.42$ & $5.72$ & $3.19$  & $5.93$ \\
\bottomrule
\end{tabular}
 
\smallskip
\footnotesize The statistic $W$ is compared with the $\chi^2_4$ distribution. The $5\%$ and $1\%$ critical values are $9.49$ and $13.28$.
\end{table}

The test rejects the null hypothesis that all clusters are weakly dependent at both levels (for the state level, at  5\% but not at 1\% level). This indicates that the variance of the cluster score grows faster than the cluster size. 

To explore the degree of dependence further, we apply the two-step procedure of Remark~\ref{rem: semi-strong} to the district-level clustered scores. We estimate $\log Z_{g,j}=\alpha_j+\beta_j\log N_g+\xi_{g,j}$ separately for each regressor with HCCM standard errors. In the first step, we test $H_0\colon\beta_j=0$ against $\beta_j>0$. In the second step, we test $H_{10}\colon\beta_j=1$ against $\beta_j<1$. Table \ref{tab:hces_loglog} reports the results.
 
\begin{table}[htbp]
\centering
\caption{Two-step procedure at the district level, HCES 2022--23}\label{tab:hces_loglog}
\begin{tabular}{lrrrrl}
\toprule
Regressor & $\hat\beta_j$ & s.e. & $p$-value ($\beta_j=0$) & $p$-value ($\beta_j=1$) & Conclusion \\
\midrule
const                   & $0.846$ & $0.111$ & $<0.001$ & $0.082$ & strong \\
$\log(\mathrm{hhsize})$ & $0.764$ & $0.092$ & $<0.001$ & $0.005$ & semi-strong \\
urban                   & $0.845$ & $0.145$ & $<0.001$ & $0.143$ & strong \\
childshare              & $0.734$ & $0.100$ & $<0.001$ & $0.004$ & semi-strong \\
\bottomrule
\end{tabular}
 
\smallskip
\footnotesize The $p$-values are one-sided, against $\beta_j>0$ and $\beta_j<1$ respectively. Conclusion uses the $5\%$ level.
\end{table}

The first step rejects $H_0\colon\beta_j=0$ for all four regressors, in line with the proposed test. The second step separates two groups. For the intercept and the urban dummy variable, $H_{10}\colon\beta_j=1$ is not rejected, so the data have strong dependence. For $\log(\mathrm{hhsize})$ and childshare, $H_{10}$ is rejected at the $1\%$ level, so the dependence is semi-strong. This two-step procedure is exploratory.

\section{Conclusion} \label{section: conclusion}

We proposed a simple test for detecting the type of within-cluster dependence in linear regression models with clustered data. The test is based on how the normalized squared cluster-level score scales with cluster size, and it is implemented through an auxiliary regression on cluster size. The test statistic is asymptotically $\chi^2_k$ under the null of weak dependence and diverges when the slope vector is non-zero under the alternative hypothesis, implying (semi) strong dependence. The covariate-specific $t$-statistics can then be used to locate the regressors along which the dependence arises. In the intercept-only model, the type of dependence is detected directly from the error covariance matrix. Simulation results show that the tests have desirable size, and power rises with the number of clusters. The empirical illustration detects non-weak dependence at the state and district levels. The test gives applied researchers a diagnostic for whether the usual large-$G$ asymptotics for cluster-robust inference are plausible. 

The test relies on variation in cluster sizes and cannot be applied when all clusters have the same size, for example, in balanced panel data. However, for unbalanced, or artificially constructed unbalanced panel, this test seems to be applicable. Future research may commence in this direction. Extensions to multi-way clustering are left for future work. 

\appendix
\section{Proofs} \label{appendix: proofs}

By $\|\cdot\|$ of a matrix, we mean the spectral norm.

Throughout, since $N_g$ is non-stochastic, $\Sigma_g$ and $V_G$ are deterministic, and
\begin{equation}\label{eq: fourth power bound}
\sum_{g=1}^G(N_g-\bar N)^4\le\max_{1\le g\le G}(N_g-\bar N)^2\sum_{g=1}^G(N_g-\bar N)^2\le\Big(\sum_{g=1}^G(N_g-\bar N)^2\Big)^2 .
\end{equation}
Under Assumption~\ref{assump: auxiliary error moments with Xg}, for every unit vector $\lambda$, $\lambda'\Sigma_g\lambda\le\sqrt{\mathbb{E}[(\lambda'\eta_g)^4]}\le\sqrt\kappa$. Hence, $\mathrm{tr}(\Sigma_g)\le k\sqrt\kappa$ and
\begin{equation}\label{eq: V_G bounds}
\sigma_{\min}^2\sum_{g=1}^G(N_g-\bar N)^2\,I_k\preceq V_G\preceq\sqrt\kappa\sum_{g=1}^G(N_g-\bar N)^2\,I_k .
\end{equation}

\begin{proof}[\textbf{Proof of Lemma \ref{lemma: bias}}]
Since $\hat\theta-\theta=\big(\sum_{h=1}^GX_h'X_h\big)^{-1}\sum_{h=1}^GX_h'\epsilon_h$,
\[
X_{g,j}'e_g^{*} = X_{g,j}'\epsilon_g - X_{g,j}'X_g\Big(\sum_{h=1}^GX_h'X_h\Big)^{-1}\sum_{h=1}^GX_h'\epsilon_h .
\]
Conditionally on $X=(X_1,\dots,X_G)$, by Assumption \ref{assump: error independence and exogenous}, we have
\begin{align*}
\mathbb{E}\Big[X_{g,j}'\epsilon_g\Big(\sum_{h=1}^GX_h'\epsilon_h\Big)'\,\Big|\,X\Big]&=X_{g,j}'\Omega_gX_g,\\
\mathbb{E}\Big[\Big(\sum_{h=1}^GX_h'\epsilon_h\Big)\Big(\sum_{h=1}^GX_h'\epsilon_h\Big)'\,\Big|\,X\Big]&=\sum_{h=1}^G X_h'\Omega_hX_h .
\end{align*}
Then,
\begin{align*}
\mathbb{E}[Z_{g,j}\mid X] = \frac{1}{N_g}\bigg[&X_{g,j}'\Omega_gX_{g,j} - 2X_{g,j}'\Omega_gX_g\Big(\sum_{h=1}^GX_h'X_h\Big)^{-1}X_g'X_{g,j}\\
&+ X_{g,j}'X_g\Big(\sum_{h=1}^GX_h'X_h\Big)^{-1}\Big(\sum_{h=1}^GX_h'\Omega_hX_h\Big)\Big(\sum_{h=1}^GX_h'X_h\Big)^{-1}X_g'X_{g,j}\bigg].
\end{align*}
By Assumption~\ref{assump: design restrictions}, $\lambda_{\min}\big(\sum_{h=1}^GX_h'X_h\big)\ge c_xN_{tot}$, so that
\[
\Big\|\Big(\sum_{h=1}^GX_h'X_h\Big)^{-1}\Big\|=\frac{1}{\lambda_{\min}\big(\sum_{h=1}^GX_h'X_h\big)}\le\frac{1}{c_xN_{tot}}.
\]
Also, $\|X_h\|^2\le C_xN_h$ for all $h$, $\|X_{g,j}\|^2\le C_xN_g$ and $\|X_g'X_{g,j}\|\le C_xN_g$. Under weak dependence, by Definition~\ref{def: dependence}, there exists a constant $C>0$ such that $\|\Omega_h\|=\lambda_{\max}(\Omega_h)\le C$ for all $h$. Hence,
\begin{align*}
\Big\|\sum_{h=1}^GX_h'\Omega_hX_h\Big\|
&\le\sum_{h=1}^G\|X_h'\|\,\|\Omega_h\|\,\|X_h\|
=\sum_{h=1}^G\|\Omega_h\|\,\|X_h\|^2
\le C\,C_x\sum_{h=1}^GN_h
=C\,C_x\,N_{tot}.
\end{align*}
For the second term,
\begin{align*}
\bigg|X_{g,j}'\Omega_gX_g\Big(\sum_{h=1}^GX_h'X_h\Big)^{-1}X_g'X_{g,j}\bigg|
&\le\|X_{g,j}\|\,\|\Omega_g\|\,\|X_g\|\,\Big\|\Big(\sum_{h=1}^GX_h'X_h\Big)^{-1}\Big\|\,\|X_g'X_{g,j}\|\\
&\le\sqrt{C_xN_g}\cdot C\cdot\sqrt{C_xN_g}\cdot\frac{1}{c_xN_{tot}}\cdot C_xN_g
=\frac{C\,C_x^2}{c_x}\,\frac{N_g^2}{N_{tot}}.
\end{align*}
For the third term,
\begin{align*}
&\bigg|X_{g,j}'X_g\Big(\sum_{h=1}^GX_h'X_h\Big)^{-1}\Big(\sum_{h=1}^GX_h'\Omega_hX_h\Big)\Big(\sum_{h=1}^GX_h'X_h\Big)^{-1}X_g'X_{g,j}\bigg|\\
&\qquad\le\|X_g'X_{g,j}\|^2\,\Big\|\Big(\sum_{h=1}^GX_h'X_h\Big)^{-1}\Big\|^2\,\Big\|\sum_{h=1}^GX_h'\Omega_hX_h\Big\|\\
&\qquad\le C_x^2N_g^2\cdot\frac{1}{c_x^2N_{tot}^2}\cdot C\,C_x\,N_{tot}
=\frac{C\,C_x^3}{c_x^2}\,\frac{N_g^2}{N_{tot}}.
\end{align*}
Dividing by $N_g$ and adding the bounds,
\begin{equation}\label{eq: explicit bias bound}
\left|\mathbb{E}[Z_{g,j}\mid X]-\frac{X_{g,j}'\Omega_gX_{g,j}}{N_g}\right|
\le\Big(\frac{2\,C\,C_x^2}{c_x}+\frac{C\,C_x^3}{c_x^2}\Big)\frac{N_g}{N_{tot}}
\le\Big(\frac{2\,C\,C_x^2}{c_x}+\frac{C\,C_x^3}{c_x^2}\Big)\max_{1\le g\le G}\frac{N_g}{N_{tot}},
\end{equation}
which tends to zero uniformly in $g$ by Assumption~\ref{assump: non dominance condition}.
\end{proof}

\setcounter{lemma}{0}
\renewcommand{\thelemma}{A.\arabic{lemma}}

\begin{lemma} \label{lemma: CLT}
Under the assumptions of Theorem \ref{thm: wald}, as $G\to\infty$,
\[
V_G^{-1/2}S_G\;\xrightarrow{d}\;N(0,I_k),\qquad \text{ where } \; V_G=\sum_{g=1}^G(N_g-\bar N)^2\Sigma_g .
\]
\end{lemma}

\begin{proof}[\textbf{Proof of Lemma \ref{lemma: CLT}}]
Under $H_0$, by \eqref{eq: S_G decomposition}, $S_G=\sum_{g=1}^G(N_g-\bar N)\eta_g$. Fix a unit vector $\lambda\in\mathbb R^k$ and write
\[
\lambda'V_G^{-1/2}S_G=\sum_{g=1}^G\xi_g,\qquad \xi_g=(N_g-\bar N)\,\lambda'V_G^{-1/2}\eta_g .
\]
By Assumption~\ref{assump: auxiliary error mean 0 with Xg}, the $\xi_g$ are independent with mean zero. Note that
\[
\sum_{g=1}^G\mathrm{Var}(\xi_g)=\lambda'V_G^{-1/2}V_GV_G^{-1/2}\lambda=1 .
\]
To show $\sum_{g=1}^G\xi_g\xrightarrow{d}N(0,1)$, we use the Lyapunov CLT with $\delta=2$. Therefore, it is required to show that the Lyapunov ratio vanishes, i.e.,
\begin{equation}\label{eq: Lyapunov ratio}
\frac{\sum_{g=1}^G\mathbb{E}[\xi_g^4]}{\left(\sum_{g=1}^G\mathrm{Var}(\xi_g)\right)^2}=\sum_{g=1}^G\mathbb{E}[\xi_g^4]\to0 .
\end{equation}
Since $V_G$ is positive definite, $V_G^{-1/2}\lambda\neq0$, and $V_G^{-1/2}\lambda\big/\|V_G^{-1/2}\lambda\|$ is a non-random unit vector. From Assumption~\ref{assump: auxiliary error moments with Xg}, we have
\[
\mathbb{E}\Bigg[\Bigg(\Big(\tfrac{V_G^{-1/2}\lambda}{\|V_G^{-1/2}\lambda\|}\Big)'\eta_g\Bigg)^4\Bigg]\le\kappa.
\]
Therefore, using $\|V_G^{-1/2}\lambda\|^2\le\big(\sigma_{\min}^2\sum_{h=1}^G(N_h-\bar N)^2\big)^{-1}$ by \eqref{eq: V_G bounds},
\begin{align*}
    \mathbb{E}[\xi_g^4] & =(N_g-\bar N)^4\,\mathbb{E}\big[(\lambda'V_G^{-1/2}\eta_g)^4\big] \\
    & = (N_g-\bar N)^4\,\big\|V_G^{-1/2}\lambda\big\|^4\;\mathbb{E}\left[\Bigg(\Big(\tfrac{V_G^{-1/2}\lambda}{\|V_G^{-1/2}\lambda\|}\Big)'\eta_g\Bigg)^4\right] \\
    & \leq \frac{\kappa\,(N_g-\bar N)^4}{\sigma_{\min}^4\big(\sum_{h=1}^G(N_h-\bar N)^2\big)^2}.
\end{align*}
Summing over $g$ and using \eqref{eq: fourth power bound},
\[
\sum_{g=1}^G\mathbb{E}[\xi_g^4]\le\frac{\kappa}{\sigma_{\min}^4}\,\frac{\sum_{g=1}^G(N_g-\bar N)^4}{\big(\sum_{g=1}^G(N_g-\bar N)^2\big)^2}\le\frac{\kappa}{\sigma_{\min}^4}\,\frac{\max_{1\le g\le G}(N_g-\bar N)^2}{\sum_{g=1}^G(N_g-\bar N)^2}\to0
\]
by Assumption~\ref{assump: no dominance variation}. Hence, \eqref{eq: Lyapunov ratio} holds, and $\lambda'V_G^{-1/2}S_G\xrightarrow{d}N(0,1)$ for every unit vector $\lambda$. By the Cram\'er--Wold device, $V_G^{-1/2}S_G\xrightarrow{d}N(0,I_k)$.
\end{proof}

\begin{lemma} \label{lemma: variance}
Under the assumptions of Theorem \ref{thm: wald}, as $G\to\infty$,
\[
\big\|V_G^{-1/2}\hat V_GV_G^{-1/2}-I_k\big\|\xrightarrow{p}0 .
\]
\end{lemma}

\begin{proof}[\textbf{Proof of Lemma \ref{lemma: variance}}]
Let $\tilde V_G=\sum_{g=1}^G(N_g-\bar N)^2\eta_g\eta_g'$, so that $\mathbb{E}[\tilde V_G]=V_G$.

\medskip
\noindent\emph{Step 1:} In this step, we show that $\|\tilde V_G-V_G\|=o_P\big(\sum_{g=1}^G(N_g-\bar N)^2\big)$.

Using the independence across clusters, applying the Cauchy--Schwarz inequality, and Assumption~\ref{assump: auxiliary error moments with Xg}, for each $j,j_1 = 1,2,\dots, k$, we have
\begin{align*}
\mathrm{Var}\big(\tilde V_{G(j,j_1)}\big)
& =\sum_{g=1}^G(N_g-\bar N)^4\,\mathrm{Var}\big(\eta_{g,j}\,\eta_{g,j_1}\big)\\
&\le\sum_{g=1}^G(N_g-\bar N)^4\,\mathbb{E}\big[\eta_{g,j}^2\,\eta_{g,j_1}^2\big]\\
&\le\sum_{g=1}^G(N_g-\bar N)^4\,\sqrt{\mathbb{E}\big[\eta_{g,j}^4\big]\;\mathbb{E}\big[\eta_{g,j_1}^4\big]}\\
&\le\sum_{g=1}^G(N_g-\bar N)^4\,\sqrt{\kappa\cdot\kappa}\\
&=\kappa\sum_{g=1}^G(N_g-\bar N)^4 .
\end{align*}
Since the spectral norm is at most the Frobenius norm,
\[
\mathbb{E}\big\|\tilde V_G-V_G\big\|^2
\le\mathbb{E}\left[\sum_{j=1}^k\sum_{j_1=1}^k\big(\tilde V_{G(j,j_1)}-V_{G(j,j_1)}\big)^2\right] =\sum_{j=1}^k\sum_{j_1=1}^k\mathrm{Var}\big(\tilde V_{G(j,j_1)}\big).
\]
Then, by \eqref{eq: fourth power bound}, we have
\[
\frac{\mathbb{E}\|\tilde V_G-V_G\|^2}{\left(\sum_{g=1}^G(N_g-\bar N)^2\right)^2}\leq \frac{k^2\kappa\sum_{g=1}^G(N_g-\bar N)^4}{\left(\sum_{g=1}^G(N_g-\bar N)^2\right)^2} \le k^2\kappa\,\frac{\max_{1\le g\le G}(N_g-\bar N)^2}{\sum_{g=1}^G(N_g-\bar N)^2}\to0
\]
by Assumption~\ref{assump: no dominance variation}. Hence, $\|\tilde V_G-V_G\|=o_P\big(\sum_{g=1}^G(N_g-\bar N)^2\big)$.

\medskip
\noindent\emph{Step 2:} In this step, we show that $\|\hat V_G-\tilde V_G\|=o_P\big(\sum_{g=1}^G(N_g-\bar N)^2\big)$.

Here, $\hat V_G = \sum_{g=1}^G(N_g-\bar N)^2 \hat{\eta}_g \hat{\eta}_g'$. Under $H_0$, $\hat\gamma=S_G/\sum_{g=1}^G(N_g-\bar N)^2$, where, by \eqref{eq: S_G decomposition} with $\gamma=0$, $S_G=\sum_{g=1}^G (N_g-\bar N)\eta_g$. The residual is
\[
\hat\eta_g=\eta_g-\bar\eta-(N_g-\bar N)\hat\gamma,
\]
where $\bar\eta=\frac1G\sum_{h=1}^G\eta_h$. Fix a unit vector $\lambda$, and write $\lambda'\hat\eta_g=\lambda'\eta_g-c_g$ with $c_g=\lambda'\bar\eta+(N_g-\bar N)\lambda'\hat\gamma$.
Then, 
\[
\lambda'\big(\hat V_G-\tilde V_G\big)\lambda
=-2\sum_{g=1}^G(N_g-\bar N)^2\big(\lambda'\eta_g\big)c_g+\sum_{g=1}^G(N_g-\bar N)^2c_g^2 .
\]
Applying the triangle inequality and the Cauchy--Schwarz inequality,
\begin{align*}
\Big|\lambda'\big(\hat V_G-\tilde V_G\big)\lambda\Big|
&\le2\,\Bigg|\sum_{g=1}^G\Big((N_g-\bar N)\,\lambda'\eta_g\Big)\Big((N_g-\bar N)\,c_g\Big)\Bigg|+\sum_{g=1}^G(N_g-\bar N)^2c_g^2\\
&\le2\,\Bigg(\sum_{g=1}^G(N_g-\bar N)^2\big(\lambda'\eta_g\big)^2\Bigg)^{1/2}\Bigg(\sum_{g=1}^G(N_g-\bar N)^2c_g^2\Bigg)^{1/2}+\sum_{g=1}^G(N_g-\bar N)^2c_g^2\\
&=2\Big(\lambda'\tilde V_G\lambda\Big)^{1/2}\Bigg(\sum_{g=1}^G(N_g-\bar N)^2c_g^2\Bigg)^{1/2}+\sum_{g=1}^G(N_g-\bar N)^2c_g^2.
\end{align*}
Since $c_g^2\le2(\lambda'\bar\eta)^2+2(N_g-\bar N)^2(\lambda'\hat\gamma)^2$,
\begin{equation} \label{eq: cros-term with c_g^2 expansion}
    \sum_{g=1}^G(N_g-\bar N)^2c_g^2\le2(\lambda'\bar\eta)^2\sum_{g=1}^G(N_g-\bar N)^2+2(\lambda'\hat\gamma)^2\sum_{g=1}^G(N_g-\bar N)^4 .
\end{equation}

Now, note that $\mathbb{E}[(\lambda'\bar\eta)^2] = G^{-2}\sum_{g=1}^G\lambda'\Sigma_g\lambda\le \sqrt\kappa/G$ by \eqref{eq: V_G bounds}. Hence, the first term in \eqref{eq: cros-term with c_g^2 expansion} is $O_P\big(\sum_{g=1}^G(N_g-\bar N)^2/G\big)=o_P\big(\sum_{g=1}^G(N_g-\bar N)^2\big)$.

Also, $\mathbb{E}[(\lambda'\hat\gamma)^2] = \lambda' V_G\lambda\big/\big(\sum_{g=1}^G(N_g-\bar N)^2\big)^2\le \sqrt\kappa\big/\sum_{g=1}^G(N_g-\bar N)^2$ by \eqref{eq: V_G bounds}. Hence, by \eqref{eq: fourth power bound}, the second term in \eqref{eq: cros-term with c_g^2 expansion} has expectation at most $2\sqrt\kappa\,\max_g(N_g-\bar N)^2$, which is $o\big(\sum_{g=1}^G(N_g-\bar N)^2\big)$ by Assumption~\ref{assump: no dominance variation}.

Hence, $\lambda'(\hat V_G-\tilde V_G)\lambda=o_P\big(\sum_{g=1}^G(N_g-\bar N)^2\big)$ for each fixed unit vector $\lambda$. Since $k$ is fixed, $\|\hat V_G-\tilde V_G\|=o_P\big(\sum_{g=1}^G(N_g-\bar N)^2\big)$.

\medskip
Steps 1 and 2 give $\|\hat V_G-V_G\|=o_P\big(\sum_{g=1}^G(N_g-\bar N)^2\big)$ by the triangle inequality. By \eqref{eq: V_G bounds},
\[
\big\|V_G^{-1/2}\big\|^2 =\frac{1}{\lambda_{\min}(V_G)}\le\frac{1}{\sigma_{\min}^2\sum_{g=1}^G(N_g-\bar N)^2}.
\]
Using the norm inequality for multiplication,
\begin{align*}
\big\|V_G^{-1/2}\hat V_GV_G^{-1/2}-I_k\big\|
&=\big\|V_G^{-1/2}\big(\hat V_G-V_G\big)V_G^{-1/2}\big\|\\
&\le\big\|V_G^{-1/2}\big\|^2\,\big\|\hat V_G-V_G\big\|\\
&\le\frac{\|\hat V_G-V_G\|}{\sigma_{\min}^2\sum_{g=1}^G(N_g-\bar N)^2}\\
&\xrightarrow{p}0 .
\end{align*}
\end{proof}

\begin{proof}[\textbf{Proof of Theorem \ref{thm: wald}}]

Write
\[
W=S_G'\hat V_G^{-1}S_G=\big(V_G^{-1/2}S_G\big)'\big(V_G^{-1/2}\hat V_GV_G^{-1/2}\big)^{-1}\big(V_G^{-1/2}S_G\big).
\]
By Lemma~\ref{lemma: variance}, we have $\big\|V_G^{-1/2}\hat V_GV_G^{-1/2}-I_k\big\|\xrightarrow{p}0$, that is, $V_G^{-1/2}\hat V_GV_G^{-1/2}\xrightarrow{p}I_k$. Since $V_G^{-1/2}\hat V_GV_G^{-1/2}\xrightarrow{p}I_k$, its eigenvalues converge in probability to one, so with probability tending to one this matrix is invertible. Therefore, $\hat V_G=V_G^{1/2}\big(V_G^{-1/2}\hat V_GV_G^{-1/2}\big)V_G^{1/2}$ is invertible with probability tending to one. By the continuous mapping theorem, the inverse of $V_G^{-1/2}\hat V_GV_G^{-1/2}$ also converges in probability to $I_k$. By Lemma~\ref{lemma: CLT}, $V_G^{-1/2}S_G\xrightarrow{d}N(0,I_k)$, under $H_0$. Hence, applying the Slutsky's theorem and the continuous mapping theorem, $W\xrightarrow{d}\chi^2_{(k)}$, under $H_0$.
\end{proof}

\begin{proof}[\textbf{Proof of Proposition \ref{prop:consistency_mult}}]

To show that $W=S_G'\hat V_G^{-1}S_G$ diverges in probability, first, note that under $H_1$, by \eqref{eq: S_G decomposition}, 
\begin{equation} \label{eq: S_G expression under H1}
    S_G=\sum_{g=1}^G(N_g-\bar N)Z_g=\gamma\sum_{g=1}^G(N_g-\bar N)^2+\sum_{g=1}^G(N_g-\bar N)\eta_g.
\end{equation}
We assume that $\hat V_G$ is invertible with probability approaching one, and $\lambda'V_G\lambda>0$ for $\lambda=\gamma/\|\gamma\|$. The Cauchy--Schwarz inequality $(\lambda'S_G)^2\le(\lambda'\hat V_G\lambda)(S_G'\hat V_G^{-1}S_G)$ gives
\begin{equation} \label{eq: W lower bound}
    W\ge\frac{(\lambda'S_G)^2}{\lambda'\hat V_G\lambda}.
\end{equation}
So, it suffices to show that ${|\lambda'S_G|} \big/{\sqrt{\lambda'\hat V_G\lambda}}$ diverges in probability.
Dividing the numerator and denominator of this ratio by $\sqrt{\lambda'V_G\lambda}$, from \eqref{eq: S_G expression under H1}, we have
\begin{equation}\label{eq: t ratio under H1}
    \frac{|\lambda'S_G|}{\sqrt{\lambda'\hat V_G\lambda}}=\frac{\Big|\;\|\gamma\|\sum_{g=1}^G(N_g-\bar N)^2\big/\sqrt{\lambda'V_G\lambda}\;+\;\lambda'\sum_{g=1}^G(N_g-\bar N)\eta_g\big/\sqrt{\lambda'V_G\lambda}\;\Big|}{\sqrt{\lambda'\hat V_G\lambda/\lambda'V_G\lambda}} .
\end{equation}

\medskip
\noindent\emph{Step 1:} We show that the numerator in \eqref{eq: t ratio under H1} diverges in probability.

By Assumption~\ref{assump: auxiliary error mean 0 with Xg}, the second term in the numerator has mean zero and variance one. Therefore, by the Chebyshev inequality, it is bounded in probability. For the first term, $\lambda'V_G\lambda\le\max_g\lambda_{\max}(\Sigma_g)\sum_{g=1}^G(N_g-\bar N)^2$, so
\[
\frac{\|\gamma\|\sum_{g=1}^G(N_g-\bar N)^2}{\sqrt{\lambda'V_G\lambda}}\ge\|\gamma\|\sqrt{\frac{\sum_{g=1}^G(N_g-\bar N)^2}{\max_g\lambda_{\max}(\Sigma_g)}}\to\infty
\]
by Assumption~\ref{assump: Sigma_g diveregence with Xg}. Since the second term is bounded in probability while the first term tends to infinity, the numerator diverges in probability.

\medskip
\noindent\emph{Step 2:} We show that $\lambda'\hat V_G\lambda/\lambda'V_G\lambda$ is bounded in probability.

The residual for the auxiliary regression is $\lambda'\hat\eta_g=\lambda'\eta_g-\lambda'\bar\eta-(N_g-\bar N)\lambda'(\hat\gamma-\gamma)$. Applying $(a-b-c)^2\le3(a^2+b^2+c^2)$,
\begin{equation} \label{eq: VG hat scalar upper bound}
    \lambda'\hat V_G\lambda\le3\sum_{g=1}^G(N_g-\bar N)^2(\lambda'\eta_g)^2+3(\lambda'\bar\eta)^2\sum_{g=1}^G(N_g-\bar N)^2+3\big(\lambda'(\hat\gamma-\gamma)\big)^2\sum_{g=1}^G(N_g-\bar N)^4 .
\end{equation}
The first term in \eqref{eq: VG hat scalar upper bound} has expectation $3\lambda'V_G\lambda$. By independence and Assumption~\ref{assump: eta_bar variance condition with Xg}, the second term in \eqref{eq: VG hat scalar upper bound} has expectation
\[
\frac{3\sum_{g=1}^G(N_g-\bar N)^2}{G^2}\sum_{g=1}^G\lambda'\Sigma_g\lambda\le3K_1\lambda'V_G\lambda .
\]
For the third term in \eqref{eq: VG hat scalar upper bound}, $\hat\gamma=S_G\big/\sum_{g=1}^G(N_g-\bar N)^2$, and from \eqref{eq: S_G expression under H1}, we have
\[
\hat\gamma-\gamma=\frac{\sum_{g=1}^G(N_g-\bar N)\eta_g}{\sum_{g=1}^G(N_g-\bar N)^2}.
\]
Since the $\eta_g$ are independent with mean zero, 
\begin{align*}
\mathbb{E}\Big[\big(\lambda'(\hat\gamma-\gamma)\big)^2\Big]
&=\frac{\sum_{g=1}^G(N_g-\bar N)^2\lambda'\Sigma_g\lambda}{\Big(\sum_{g=1}^G(N_g-\bar N)^2\Big)^2}
=\frac{\lambda'V_G\lambda}{\Big(\sum_{g=1}^G(N_g-\bar N)^2\Big)^2}.
\end{align*}
Using $\sum_{g=1}^G(N_g-\bar N)^4\le\big(\sum_{g=1}^G(N_g-\bar N)^2\big)^2$, the third term in \eqref{eq: VG hat scalar upper bound} has expectation at most $3\lambda'V_G\lambda$.

Therefore, $\mathbb{E}[\lambda'\hat V_G\lambda]\le(6+3K_1)\lambda'V_G\lambda$, and by Markov's inequality, for every $K>0$,
\[
\mathbb{P}\Big(\frac{\lambda'\hat V_G\lambda}{\lambda'V_G\lambda}>K^2\Big)\le\frac{6+3K_1}{K^2}.
\]
Since the right-hand side tends to zero as $K\to\infty$, $\lambda'\hat V_G\lambda/\lambda'V_G\lambda$ is bounded in probability.

By Step 1, the numerator in \eqref{eq: t ratio under H1} diverges in probability, and by Step 2, the square of its denominator is bounded in probability. Hence, the ratio in \eqref{eq: t ratio under H1} diverges in probability. Therefore, $\mathbb{P}(W>\delta)\to1$ for every fixed $\delta>0$.

\end{proof}

\setcitestyle{aysep={}}
\bibliography{references}

@article{white1980heteroskedasticity,
  author  = {White, Halbert},
  title   = {A heteroskedasticity-consistent covariance matrix estimator and a direct test for heteroskedasticity},
  journal = {Econometrica},
  volume  = {48},
  number  = {4},
  pages   = {817--838},
  year    = {1980}
}

@article{moulton1990, author={Moulton, Brent R.}, title={An illustration of a pitfall in estimating the effects of aggregate variables on micro units}, journal={Review of Economics and Statistics}, year={1990}, volume={72}, pages={334--338}}

@article{cameron2015, author={Cameron, A. Colin and Miller, Douglas L.}, title={A practitioner's guide to cluster-robust inference}, journal={Journal of Human Resources}, year={2015}, volume={50}, pages={317--372}}

@article{abadie2023, author={Abadie, Alberto and Athey, Susan and Imbens, Guido W. and Wooldridge, Jeffrey M.}, title={When should you adjust standard errors for clustering?}, journal={Quarterly Journal of Economics}, year={2023}, volume={138}, pages={1--35}}

@article{carter2017, author={Carter, Andrew V. and Schnepel, Kevin T. and Steigerwald, Douglas G.}, title={Asymptotic behavior of a t-test robust to cluster heterogeneity}, journal={Review of Economics and Statistics}, year={2017}, volume={99}, pages={698--709}}

@misc{basak2018understanding, author={Basak, Gopal K. and Das, Samarjit}, title={Understanding cross-sectional dependence in panel data}, year={2018}, eprint={1804.08326}, archivePrefix={arXiv}}

@article{breusch1980lagrange,
  title={The {Lagrange} multiplier test and its applications to model specification in econometrics},
  author={Breusch, Trevor S and Pagan, Adrian R},
  journal={The Review of Economic Studies},
  volume={47},
  number={1},
  pages={239--253},
  year={1980},
  publisher={Wiley-Blackwell}
}

@article{pesaran2021general,
  title={General diagnostic tests for cross-sectional dependence in panels},
  author={Pesaran, M Hashem},
  journal={Empirical Economics},
  volume={60},
  number={1},
  pages={13--50},
  year={2021},
  publisher={Springer}
}

@article{mackinnon2023guide,
  author  = {MacKinnon, James G. and Nielsen, Morten {\O}rregaard and Webb, Matthew D.},
  title   = {Cluster-robust inference: A guide to empirical practice},
  journal = {Journal of Econometrics},
  year    = {2023},
  volume  = {232},
  pages   = {272--299}
}

@techreport{mackinnon2019when,
  author      = {MacKinnon, James G. and Webb, Matthew D.},
  title       = {When and how to deal with clustered errors in regression models},
  institution = {Queen's University, Department of Economics},
  type        = {Working Paper},
  number      = {1421},
  year        = {2019}
}

@article{djogbenou2019,
  author  = {Djogbenou, Antoine A. and MacKinnon, James G. and Nielsen, Morten {\O}rregaard},
  title   = {Asymptotic theory and wild bootstrap inference with clustered errors},
  journal = {Journal of Econometrics},
  year    = {2019},
  volume  = {212},
  pages   = {393--412}
}

@article{mackinnon2017wild,
     author  = {MacKinnon, James G. and Webb, Matthew D.},
     title   = {Wild bootstrap inference for wildly different cluster sizes},
     journal = {Journal of Applied Econometrics},
     year    = {2017},
     volume  = {32},
     pages   = {233--254}
   }

@article{hansen2019asymptotic,
	title={Asymptotic theory for clustered samples},
	author={Hansen, Bruce E and Lee, Seojeong},
	journal={Journal of econometrics},
	volume={210},
	number={2},
	pages={268--290},
	year={2019},
	publisher={Elsevier}
}

@article{moran1948interpretation,
  author  = {Moran, P. A. P.},
  title   = {The interpretation of statistical maps},
  journal = {Journal of the Royal Statistical Society: Series B},
  volume  = {10}, number = {2}, pages = {243--251}, year = {1948}
}

@book{anselin1988spatial,
  author    = {Anselin, L.},
  title     = {Spatial Econometrics: Methods and Models},
  publisher = {Kluwer Academic Publishers}, address = {Dordrecht}, year = {1988}
}

@article{pesaran2015testing,
  author  = {Pesaran, M. H.},
  title   = {Testing weak cross-sectional dependence in large panels},
  journal = {Econometric Reviews},
  volume  = {34}, pages = {1089--1117}, year = {2015}
}

@article{bailey2016exponent,
  author  = {Bailey, N. and Kapetanios, G. and Pesaran, M. H.},
  title   = {Exponent of cross-sectional dependence: Estimation and inference},
  journal = {Journal of Applied Econometrics},
  volume  = {31}, pages = {929--960}, year = {2016}
}

@article{sarafidis2012cross,
  author  = {Sarafidis, V. and Wansbeek, T.},
  title   = {Cross-sectional dependence in panel data analysis},
  journal = {Econometric Reviews},
  volume  = {31}, pages = {483--531}, year = {2012}
}

@article{pesaran2008bias,
  author  = {Pesaran, M. Hashem and Ullah, Aman and Yamagata, Takashi},
  title   = {A bias-adjusted {LM} test of error cross-section independence},
  journal = {The Econometrics Journal},
  volume  = {11},
  number  = {1},
  pages   = {105--127},
  year    = {2008}
}

@article{chudik2011weak,
  author  = {Chudik, Alexander and Pesaran, M. Hashem and Tosetti, Elisa},
  title   = {Weak and strong cross-section dependence and estimation of large panels},
  journal = {The Econometrics Journal},
  volume  = {14},
  number  = {1},
  pages   = {C45--C90},
  year    = {2011}
}

@article{wooldridge2003cluster,
	title={Cluster-sample methods in applied econometrics},
	author={Wooldridge, Jeffrey M},
	journal={American Economic Review},
	volume={93},
	number={2},
	pages={133--138},
	year={2003},
	publisher={American Economic Association}
}

@article{cameron2008bootstrap,
	title={Bootstrap-based improvements for inference with clustered errors},
	author={Cameron, A Colin and Gelbach, Jonah B and Miller, Douglas L},
	journal={The review of economics and statistics},
	volume={90},
	number={3},
	pages={414--427},
	year={2008},
	publisher={The MIT Press}
}

@article{abadie2020sampling,
  title={Sampling-based versus design-based uncertainty in regression analysis},
  author={Abadie, Alberto and Athey, Susan and Imbens, Guido W and Wooldridge, Jeffrey M},
  journal={Econometrica},
  volume={88},
  number={1},
  pages={265--296},
  year={2020},
  publisher={Wiley Online Library}
}

@article{canay2021wild,
  title={The wild bootstrap with a “small” number of “large” clusters},
  author={Canay, Ivan A and Santos, Andres and Shaikh, Azeem M},
  journal={Review of Economics and Statistics},
  volume={103},
  number={2},
  pages={346--363},
  year={2021},
  publisher={MIT Press One Rogers Street, Cambridge, MA 02142-1209, USA journals-info~…}
}

@article{chiang2023genuinely,
  title={Genuinely Robust Inference for Clustered Data},
  author={Chiang, Harold D and Sasaki, Yuya and Wang, Yulong},
  journal={arXiv preprint arXiv:2308.10138},
  year={2023}
}

@article{cameron2011robust,
  title={Robust inference with multiway clustering},
  author={Cameron, A Colin and Gelbach, Jonah B and Miller, Douglas L},
  journal={Journal of Business \& Economic Statistics},
  volume={29},
  number={2},
  pages={238--249},
  year={2011},
  publisher={Taylor \& Francis}
}

@article{mackinnon2021wild,
  title={Wild bootstrap and asymptotic inference with multiway clustering},
  author={MacKinnon, James G and Nielsen, Morten {\O}rregaard and Webb, Matthew D},
  journal={Journal of Business \& Economic Statistics},
  volume={39},
  number={2},
  pages={505--519},
  year={2021},
  publisher={Taylor \& Francis}
}

@article{chiang2022multiway,
  title={Multiway cluster robust double/debiased machine learning},
  author={Chiang, Harold D and Kato, Kengo and Ma, Yukun and Sasaki, Yuya},
  journal={Journal of Business \& Economic Statistics},
  volume={40},
  number={3},
  pages={1046--1056},
  year={2022},
  publisher={Taylor \& Francis}
}

@article{tabord2019inference,
  title={Inference with dyadic data: Asymptotic behavior of the dyadic-robust t-statistic},
  author={Tabord-Meehan, Max},
  journal={Journal of Business \& Economic Statistics},
  volume={37},
  number={4},
  pages={671--680},
  year={2019},
  publisher={Taylor \& Francis}
}

@article{kojevnikov2023some,
  title={Some impossibility results for inference with cluster dependence with large clusters},
  author={Kojevnikov, Denis and Song, Kyungchul},
  journal={Journal of Econometrics},
  volume={237},
  number={2},
  pages={105524},
  year={2023},
  publisher={Elsevier}
}

@article{mackinnon2023testing,
  title={Testing for the appropriate level of clustering in linear regression models},
  author={MacKinnon, James G and Nielsen, Morten {\O}rregaard and Webb, Matthew D},
  journal={Journal of Econometrics},
  volume={235},
  number={2},
  pages={2027--2056},
  year={2023},
  publisher={Elsevier}
}

@misc{dey2024cross,
  title={Cross Sectional Regression with Cluster Dependence: Inference based on Averaging},
  author={Dey, Subhodeep and Basak, Gopal K and Das, Samarjit},
  howpublished={arXiv preprint arXiv:2408.13514},
  year={2024}
}

@article{dey2026linear,
  title={Linear Regression: Inference Based on Cluster Estimates},
  author={Dey, Subhodeep and Basak, Gopal K and Das, Samarjit},
  journal={arXiv preprint arXiv:2602.04691},
  year={2026}
}

\end{document}